\documentclass[12pt]{iopart}

\usepackage{natbib,graphicx}
\newcommand{\aap}[2]{{\em Astr. Astrophys.}, {\bf #1}, #2}

\newcommand{\apj}[2]{{\em Astrophys. J.}, {\bf #1}, #2}

\newcommand{\apjs}[2]{{\em Astrophys. J. Supp.}, {\bf #1}, #2}

\newcommand{\pasp}[2]{{\em Pub. Astron. Soc. Pacific}, {\bf #1}, #2}
\newcommand{\pasa}[2]{{\em Pub. Astron. Soc. Australia}, {\bf #1}, #2}
\newcommand{\mnras}[2]{{\em Mon. Not. R. Astr. Soc.}, {\bf #1}, #2}

\newcommand{\nat}[2]{{\em Nature}, {\bf #1}, #2}

\def\newblock{\hskip .11em plus .33em minus .07em}

\begin{document}

\title[Techniques for Measuring Pulsar Times of Arrival]{Pulsar Timing Techniques} 

\author{Andrea N. Lommen$^1$ and Paul Demorest$^2$}

\address{$^1$ Franklin and Marshall College, Lancaster, PA}
\address{$^2$ National Radio Astronomy Observatory, Charlottesville, VA}

\ead{alommen@fandm.edu}
\begin{abstract}
We describe the procedure, nuances, issues, and choices involved in creating times-of-arrival (TOAs),
residuals and error bars from a set of radio pulsar timing data.  
We discuss the issue of mis-matched templates, the problem that wide-bandwidth backends introduce, possible
solutions to that problem, and correcting for offsets introduced by various observing systems. 
\end{abstract}

\maketitle

\section{Introduction to Pulsar Timing Arrays}
A pulsar is essentially a clock.
Pulsars pulse with stunning precision, e.g. in
the best-timed radio millisecond pulsars we can predict the arrival time of the next pulse to within 100 ns over a 5 year
period \citep{Verbiest08}.  A system of well-timed radio millisecond pulsars, therefore, is a system
of clocks, and can be used to measure spacetime disturbances such as gravitational waves.  Such as system
has been termed a ``pulsar timing array'' by \cite{Backer83} and according to current models
pulsar timing arrays will detect gravitational waves within the next decade \cite{Demorest09}.

\section{Introduction to the Problem }

The idea of pulsar timing is simple:  we time the arrival of a series of pulses, and fit 
a model to the measured arrival times.  As more pulsar data becomes available we improve the accuracy
of the model.

There is a fundamental
chicken and egg
problem in pulsar timing that goes as follows.
The pulsed radio signal is buried in the noise for most pulsars, i.e. most single pulses are undetectable.  
Thus, in order to detect a pulse, one must add together many thousands of pulses 
over many thousands of turns of the pulsar.  In order for this procedure to yield an accumulated
pulse profile with a  high signal to
noise ratio, one must have a model 
that tells you how to `fold' the pulsar data onto itself.  In order to find this model, one
must be able to detect a pulse \citep{LorimerBook12}.  The chicken and egg problem is therefore
that in order to get a model you need 
accumulated pulses, and in order to accumulate pulses you need a model.

The chicken and egg problem is solved iteratively.  When a pulsar is first discovered, the 
observer usually has a rough idea of its spin frequency by the location of the peak in the FFT
in which it was discovered.  This approximate spin frequency is used to accumulate pulses in small
batches and acquire 3 or 4 times of arrival (TOAs) for the pulsar.  A model is fit to the TOAs 
and the new model is used to fold the next set of data that comes in, and the process 
iterates.  Depending on the pulsar `backend' \footnote{`Backend' refers to the equipment
at the telescope that is used to process the voltages that come in from the `front end' or
telescope. Descriptions of backends can be found in 
\citet{Kramer99, Voute02, GUPPI08, Cognard09, Karuppusamy08, Karuppusamy11, Sarkissian11}.  
} one may be able to go back to the original data and re-fold the data using
the improved model.
Employing this iterative process results in a good timing model within a year, i.e. one that 
predicts the arrival time of the next pulse to within a few milliperiods
(0.001 times the pulse period) \citep{Verbiest09, Demorest13}.  

The iterative process continues throughout the dataspan of the pulsar, but the model is
updated much less frequently than at first, perhaps every couple of years.  For example, it is
only after a couple of years of data-taking that one is able to fit for proper motion and
parallax, so at that moment those parameters are added to the model and the model improves.

\section{What time do you say the pulse arrives?}

In the previous section we evaded the question of how we measure the
arrival time.  When a pulsar signal is folded, as described above, over
many thousands of turns, the resulting high signal-to-noise data
representing the average shape of the pulse is called a ``pulse
profile."  The shift in rotational phase between this profile and a
high signal-to-noise reference profile (usually called the
``template'' or ``standard'' profile) is then determined, typically
using a Fourier-domain algorithm \citep{Taylor92}.  We review the
standard algorithm below. 
 
\begin{figure}
\centerline{\includegraphics[width = \textwidth]{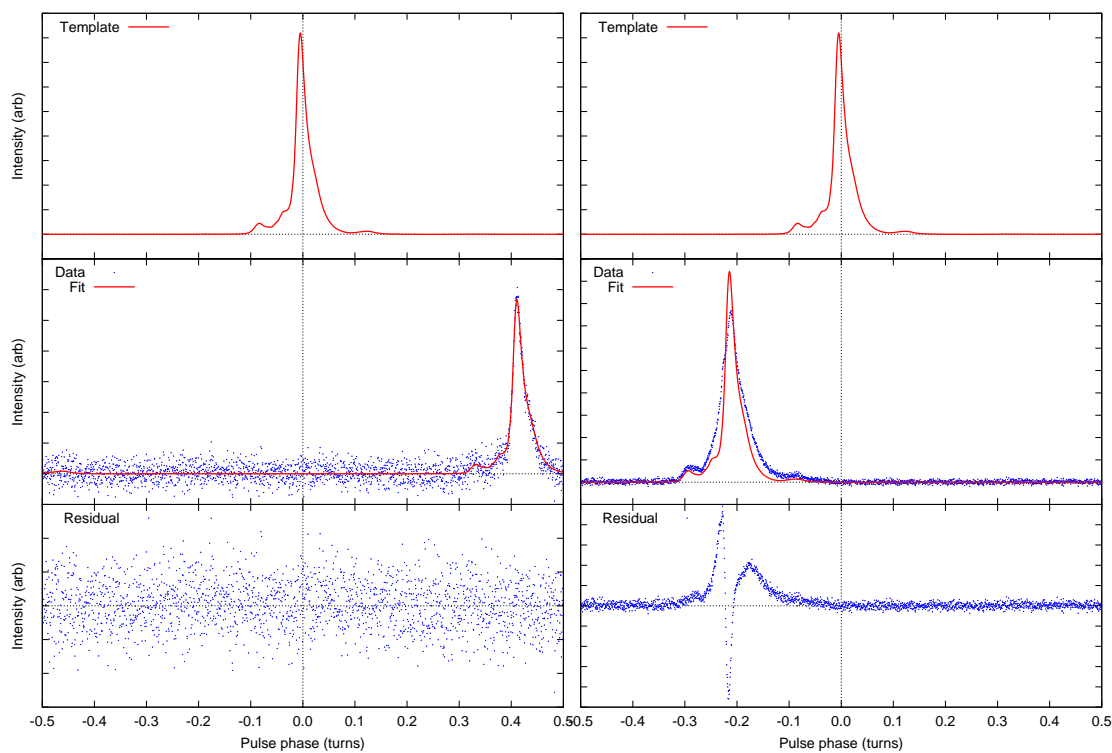}}
\caption{
Illustration of the template matching process.  The upper
panels show a template profile for the 4.6-ms pulsar J1713$+$0747
determined from 1600~MHz data taken with the Green Bank Telescope (note
the same profile is shown in both upper panels).  The fiducial point is
at phase zero, and was determined via the phase of the first Fourier
component.  In the middle left-hand panel, this template profile is fit
to a 1600~MHz ``data'' profile using the method described in the text.
The measured phase shift is $0.4158\pm0.00013$ turns.  The bottom-left
panel shows the noise-like profile residuals from this fit.  The
middle-right panel shows a fit of the same template to data taken at
750~MHz.  In this case, the template is a poor match to the data and
there is highly significant structure left over in the residuals (bottom
right).  The phase shift in this case is $0.19036\pm0.00016$ turns.
Using the correct template for the 750~MHz data results in a factor
of $\sim$3 smaller phase shift uncertainty.
\label{fig:mismatch}
}
\end{figure}

The template profile represents a model for the expected profile shape 
in the absence of noise (see Figure \ref{fig:mismatch} for an example.)  
This may be created by summing a large amount 
of profile data togther, and post-processing the result to remove 
remaining noise.  As demonstrated by \citet{Hotan05}, using a noisy 
template can produce biased results due to correlation between the noise in 
the template and the profiles from which it was generated.  Noise in the 
template is often removed by a low-pass filter.  \cite{Demorest13} used 
a wavelet-based noise removal process.  Another common technique is to create a template 
by fitting a series of Gaussian functions to a high 
signal-to-noise profile \citep{Kramer94II}. 
 
Given a data profile $d(\phi)$ and a template profile $p(\phi)$ where $\phi$ runs
from 0 to 1, the
relative phase shift between the two, $\Delta\phi$, is usually determined via a $\chi^2$
minimization in the Fourier domain:
\begin{equation}\label{eqn:chi2}
  \chi^2(a,\Delta\phi) = \sum^{k_{max}}_{k=1}
    \frac{\left|{d_k - ap_ke^{-2\pi i k \Delta\phi}}\right|^2}{\sigma^2}
\end{equation}
Here, $d_k$ and $p_k$ are the discrete Fourier transforms of the
profiles and $\sigma^2$ is the noise power in each harmonic component,
assumed to be constant.  The fit parameter $\Delta\phi$ is the phase shift of
interest, and the nuisance parameter $a$ is a scale factor between the
two profiles (related to the pulse flux)\footnote{An additional nuisance
factor -- the DC offset between the two profiles -- has been avoided by
starting the sum at $k=1$.}  It is possible to expand the sum in the
above equation into three terms:
\begin{equation}
  \chi^2(a,\Delta\phi) = \sigma^{-2}
    \left( D^2 + a^2P^2 - 2aC_{dp}(\Delta\phi) \right)
\end{equation}
Here, $D^2$ and $P^2$ are the sum of squares of the Fourier amplitudes
of the data and template profiles respectively.  The final term,
$C_{dp}(\Delta\phi) = \mathrm{Re} \sum_k d_k p_k^*e^{2\pi i k \phi}$ can be
interpreted as the cross-correlation between the two profiles.  This
term contains all the phase shift information, and it is straightforward
to show that the $\chi^2$ minimum always occurs at the phase shift that
maximizes $C_{dp}$.  We call this phase shift where $C_{dp}$ is a maximum ${\Delta\hat\phi}$.  

This phase shift represents the difference between the observed and model-predicted pulse phases, 
averaged over the timespan of the observation.  This is converted to a TOA as follows:  First, the 
model-predicted phase at the midpoint of the observation is subtracted from the measured phase shift.  
The result is multiplied by the model-predicted pulse period and added to the midpoint time, giving the 
final time of arrival.  TOAs calculated in this manner are independent of the timing model used to fold 
the data, and can then be used as input for future model fits without needing to retain full details of 
the original folding.  Referring the TOA to the midpoint of the observation (rather than the start) 
avoids biases that can be introduced due to an inaccurate initial model.  TOAs are however {\it not} 
independent of the template profile used to measure them and care must be taken when combining sets of 
TOAs determined using different templates (see \S \ref{sec:wrong} for additional discussion).

Relating the time-stamp to UTC is also a non-trivial endeavor.  The original time-stamp from the observatory is 
generally from time kept by a hydrogen maser local to the observatory.  The comparison between this ``observatory
time" and GPS ``common view" time is kept on an hourly basis and is generally smaller than 100 ns.  The comparison
between GPS time and Bureau International des Poids et Mesures (BIPM) \footnote{http://www.bipm.org} 
time is published by BIPM monthly.  These 3 ingredients are added together to yield
time stamps that are within a few nanoseconds of UTC.   Various telescopes have slight variations on this
scheme but the general idea is usually the same.  For a more complete description of this process
see \citet{Edwards06tempo2paper2}.  

An estimate of the uncertainty of the best-fit
${\Delta\hat\phi}$ (and $a$ if desired) is then found via second derivatives of
$\chi^2$ following standard procedures (e.g. \cite{NR14.5}):
\begin{equation}
  \sigma^2_{\Delta\hat\phi} = \frac{1}{2} \left(
    \frac{\partial^2 \chi^2}{\partial \phi^2} \right)^{-1} =
    \frac{\sigma^2P^2}{-2C_{dp}({\Delta\hat\phi})C^{\prime\prime}_{dp}({\Delta\hat\phi})}
\end{equation}
This uncertainty estimate implicitly assumes that the parameter
likelihood function (proportional to $e^{-\chi^2/2}$) has a Gaussian
shape.  In the high-S/N limit this is an extremely good approximation.
However, at lower S/N values, this approximation becomes worse, and the
TOA error distribution becomes non-Gaussian, and highly dependent on the
profile shape.  This is most often dealt with by employing a
signal-to-noise ratio cutoff, below which profile data is not used for
TOA generation.

The uncertainty in a TOA is roughly the pulse width $W$ divided by the signal
to noise of the profile $S/N$ and scales as follows \cite{LorimerBook12}
\begin{equation}
\sigma_\phi \approx \frac{W}{S/N} \propto \frac{S_{sys}}{\sqrt{t_{obs}\Delta f}} \times \frac{P\delta^{3/2}}{S_{mean}}
\end{equation}
where $S_{sys}$ is the flux density of the pulsar, $\Delta f$ is the observing bandwidth, $t_{obs}$ is the integration
time, $P$ is the pulse period, and $W$ is the width of the pulse, and $S_{mean}$ is the mean flux
density of the pulsar.
Bright, short-period, narrow pulsars yield smaller uncertainties as do sensitive wide-band detecting systems.
 
The uncertainty in the TOA is determined via the $\chi^2$ procedure 
described above, and depends only on the shape of the template data 
profiles, and the noise level in the data ($\sigma$).  The noise level 
is primarily due to radiometer noise, and the standard TOA uncertainty 
does not explicitly include additional sources of TOA systematic bias, 
for example from the interstellar medium (ISM) or clock errors. 
However, both ISM errors and polarization calibration errors can produce 
a mis-match between the shape of the template and the shape of the 
profile (as shown in Figure \ref{fig:mismatch})
and as such will bias TOAs by some amount, possibly in excess 
of their formal $\chi^2$-estimated uncertainties.  
 
Many pulsars exhibit evidence for additional sources of noise beyond the 
simple radiometer-noise-dominated measurement errors(\citet{Hobbs06}, also see articles 
by Stinebring and Cordes in this volume) and it is less clear 
how to account for these in the estimate of uncertainty in the TOAs. First, 
the pulsar is likely to have intrinsic phase or frequency noise (see article by Cordes in this volume and
also \citep{Shannon10}) but it is unclear 
at what level this enters. 
In either 
case this intrinsic noise is not easy to estimate, and is not generally reflected in the formally calculated
uncertainties. 
Similarly, the interstellar medium (ISM) adds noise both by changing the index of refraction along the line-of-sight to 
the pulsar, and by scattering the pulse (see article by Stinebring in this issue).  Both effects cause delays.  The standard 
way of accounting for the former is to take data at two radio frequencies separated substantially in frequency 
(such as 1400 and 2400 MHz, see e.g. \citep{Kaspi94}).  The difference in arrival time between the two frequencies allows one
 to calculate the 
``dispersion measure" or DM
along the line of sight and allows the subtraction of the delay caused by the ISM as follows:   
\begin{eqnarray}
t_2 - t_1 = (1/0.241)~{\rm ms~DM} \left[\left(\nu_1/{\rm GHz}\right)^{-2} - \left(\nu_2/{\rm GHz}\right)^{-2}\right]
\end{eqnarray}
where $t_1$ and $t_2$ are the arrival times at the two frequencies $\nu_1$ and $\nu_2$ and DM is
proportional the column density of electrons along the line of site\citep{LorimerBook12}.
With current instruments, variation of DM over time is detectable in the timing of most bright MSPs, and correcting 
them via a technique such as this is necessary to achieve $\sim$~100-ns level timing (\citep{Demorest13, 
Keith13}). Recent work by \citet{Shannon10} suggests that 3- or 4-frequency data and more 
advance corrections schemes may be necessary to achieve sub-100-ns results for many pulsars.
However, the newest pulsar machines (see articles 
by McLaughlin, Kramer and Hobbs in this volume) have such wide bandwidths ($\sim$ 1 GHz) that it may be 
possible to leverage the bandwidth to accomplish the same thing.  This is a bit tricky, as we will describe in the 
next section.  \citet{Keith13} found success using the same equation above in conjunction with a
``common mode" TOA and an iterative process to find DM(t).  

In addition, pulsar backends must deal with dispersion within (not just between) observing bands.  For a full treatment
of this issue please see \citet{LorimerBook12}.
 
One common software package used for measuring TOAs (as well as 
calibration and a large number of other standard data processing steps) 
is called PSRCHIVE\footnote{http://psrchive.sourceforge.net} 
\citep{Hotan04, vanStraten12}.  PSRCHIVE is developed collaboratively 
by members of the pulsar research community worldwide, and is freely 
available, open-source software. 

\section{Fitting to a model}
\label{sec:fitting}

The basic ingredients of a pulsar timing model are the spin period and period derivative,
right ascension (RA) and declination (DEC) of a pulsar.
With these ingredients one can usually construct a phase-connected solution that spans several months, or perhaps even
a year.  If the pulsar is in a binary system, one needs 5 additional Keplerian orbital parameters:  the projected
semi-major axis of the orbit, the orbital period, the position of peri-astron, the eccentricity, and the mass function.
Other circumstances and additional refinements
can warrant additional parameters:  relativistic orbits, glitches, proper motion, parallax, etc \citep{LorimerBook12}.
The criterion of ``phase-connection" refers to the necessity for the model to keep track of which turn the pulsar is
on.  For example a particular model may do well to predict the arrival time of several pulses separated by a couple days, but when
another data point is added a month later, the model may not be good enough to predict the 4th point and connect its phase.  In fact if the model
is off by more than a pulse-period, the model will not even be able to correctly calculate which turn the
pulsar is on (much less its phase), and
therefore the fit to such a model produces erratic results.  A phase-connected model will correctly predict the arrival time
of every pulse corresponding to a TOA to within a half of a pulse period.  For more information on the details
of timing model fits, see Edwards, Hobbs, \& Manchester (2006).\nocite{Edwards06tempo2paper2}.

\begin{figure}
\includegraphics[width=0.5\textwidth]{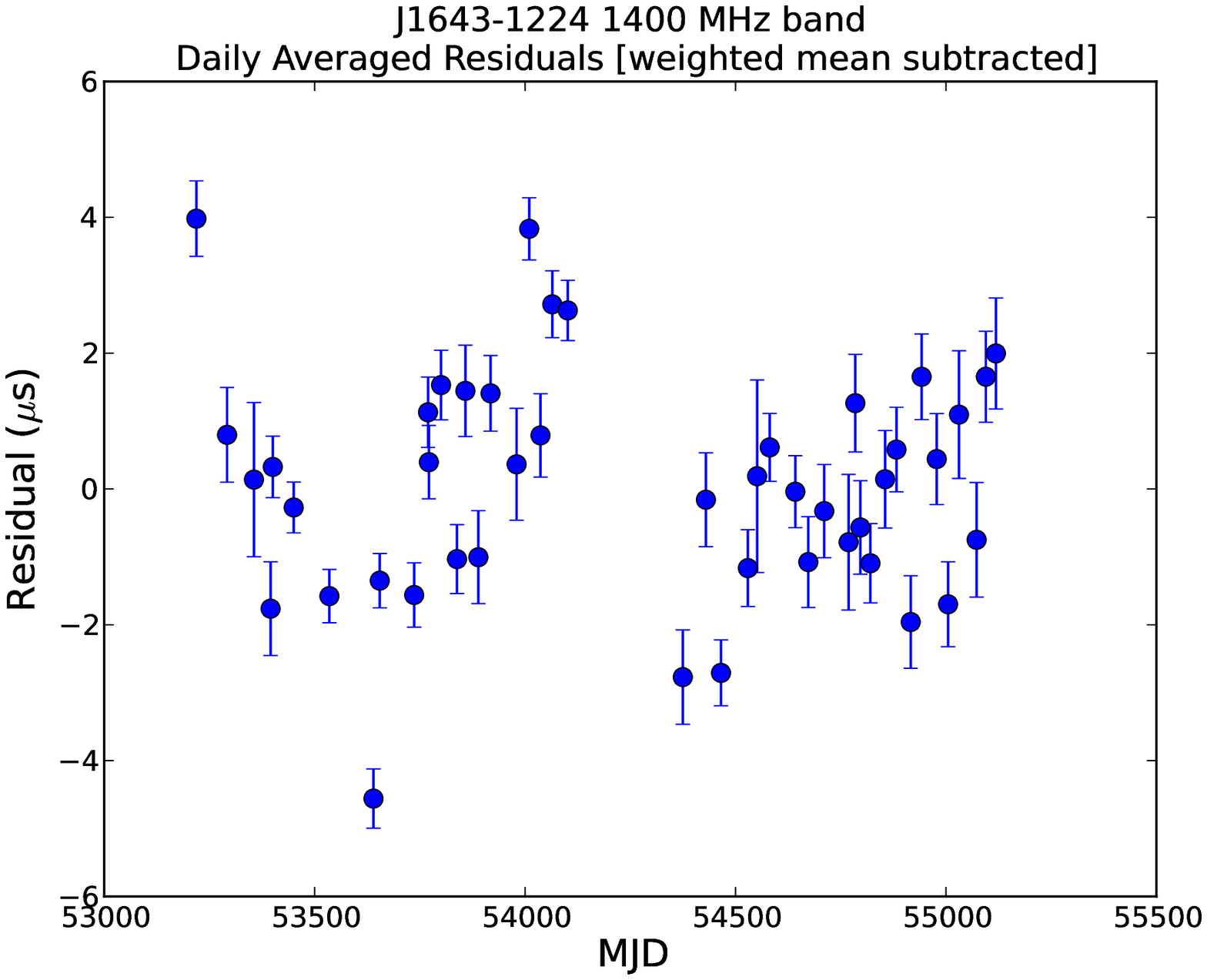}
\includegraphics[width=0.5\textwidth]{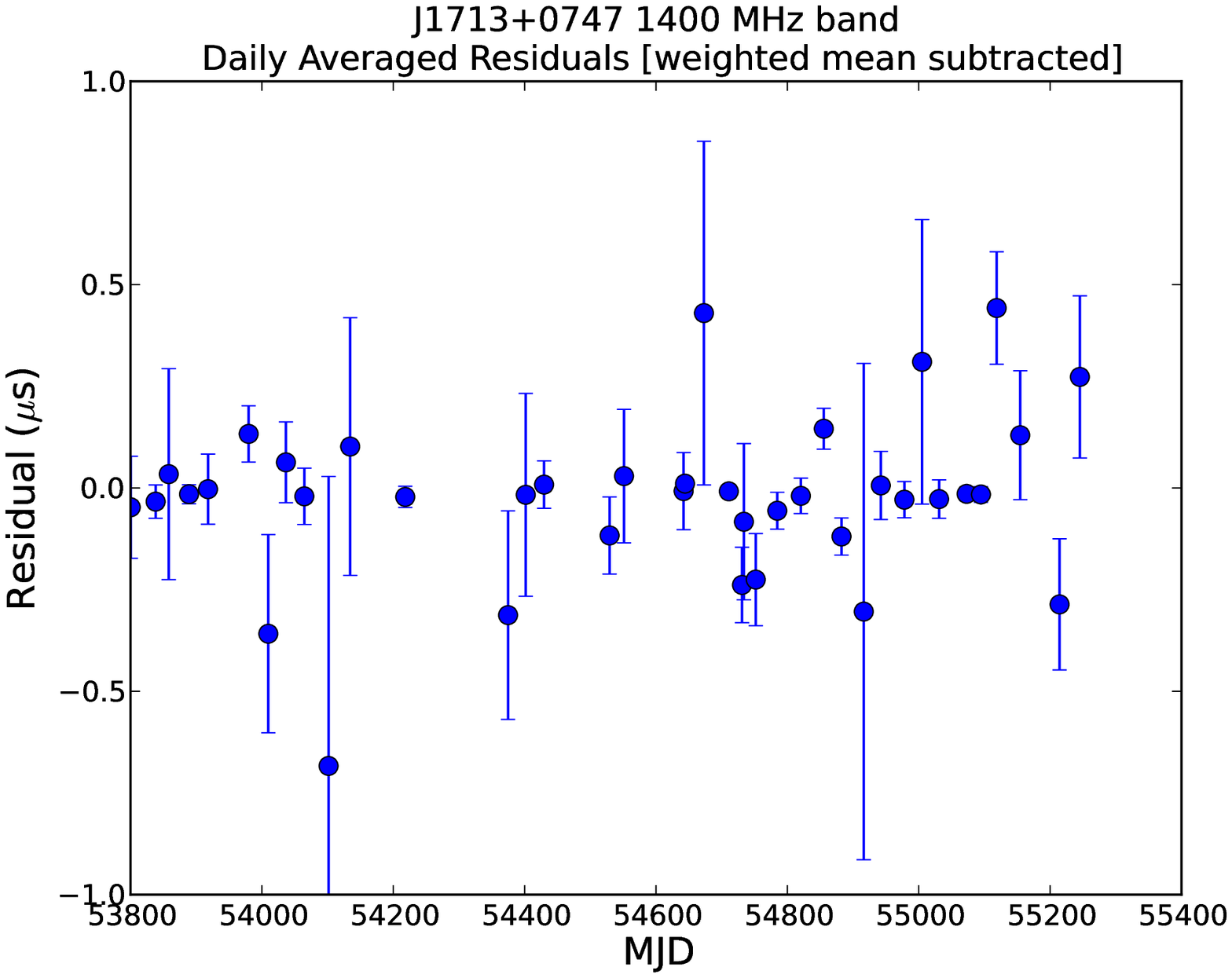}
\caption{Residuals from PSR J1643$-$1224 (left panel), PSR J1713$+$0747 (right panel).  On the left an example of pulsar timing residuals where the scatter is larger than the error bars warrant, and on the right an example where the error bars are consistent with zero residual.  Data are publicly available North American Nanohertz Observatory of Gravitational Wave (NANOGrav) data from the Arecibo Observatory and the Green Bank Telescope\citep{Demorest13}.
\label{fig:residuals}
}
\end{figure}

The output of the model-fitting is a list of the fitted parameters and their uncertainties, but also the so-called
``residuals," i.e. the difference between the model and the measured TOAs.  For high signal-to-noise data a 
good model generally yields residuals
that are in the vicinity of or better than a milliperiod.
Two example plots of residuals are shown in Figure \ref{fig:residuals}.   

The fit of a model to the TOAs is done using the more traditional TEMPO\footnote{http://tempo.sourceforge.net} or the
newer TEMPO2\footnote{http://www.sf.net/projects/tempo2/}.  Both are freely available for download, contain
most of the same features, and for
the remainder of this article both will be implied when we say ``TEMPO."  In both cases
the input to the program is a list of TOAs (produced as described above) plus a timing model that fits the data well enough
for phase connection as described above.  If the data are indeed phase connected then the least-squares fitting algorithm
performed in TEMPO will improve the fit (as judged by its $\chi^2$ or by its RMS residual) and refine the fitted
parameters.  If the model is not sufficiently good to phase connect the timing solution then the fitting procedure will
not achieve anything worthwhile, and is likely to find a solution in local minimum that is even worse than the original input model.

Fits often yield large $\chi^2$ values indicating that the error bars on the timing points have been underestimated or there
is unmodeled noise, perhaps
due to some of the issues discussed in the previous section.
Pulsar timers commonly multiply the errors acquired from the comparison of the template to the profile (using the  
parameter EFAC in TEMPO, software described in section \ref{sec:fitting}) or add in a quadrature another component (using the parameter EQUAD).  This 
is a practice that is empirical, based on noticing that the scatter of residuals exceeds what you would 
expect from the size of the error bars (such as in the left-hand panel of Figure \ref{fig:residuals}).   
 
The practice of adding arbitrarily to the error bars 
 is admittedly unappealing; there are other solutions. 
One method involves using the scatter of the residuals on a particular day, typically the standard 
deviation of them, as an estimate of the uncertainty from 
these unknown sources.  This accounts for noise on shorter timescales (typically data are taken over 
a half-hour period) but does not account for noise on longer timescales (many of the timescales in 
the interstellar medium are multiple days.)  
 
Large $\chi^2$ values in the timing fit also can be indicative of red noise in the residuals.
Both \citet{Coles11} and \citet{vanHaasteren13} have 
developed advanced techniques for dealing with correlated noise in pulsar timing residuals.  \citet{Coles11} employs a whitening procedure 
before fitting, and \citet{vanHaasteren13} develops a Bayesian technique for parameter estimation in the presence of red noise.

One way to improve this procedure is to find values of fitted parameters in ways other than performing timing fits, such as finding
the RA and DEC and proper motion of the pulsar using very-long baseline interferometry (VLBI).  This has advantages.  
First, it can allow one to find
a phase-connected solution earlier.  Second, it will reduce the uncertainties in parameters that covary with the astrometric
parameters (especially in shorter data sets ($<~1$ year), the pulsar parameters (e.g. period derivative)
 covary with the astrometric parameters (e.g. right ascension).  This elimination of covariance has favorable
consequences such as allowing pulsars to be assessed for the presence of red noise
more accurately earlier than without \citep{Madison12}.  \citet{Madison12} also showed that a sub-milliarcsec determination of a pulsar's
position and proper motion using VLBI would allow some gravitational wave signatures to be found more easily.

\section{Daily averaging of TOAs}


Commonly, the data from the telescope arrives in one-minute ``scans" and a typical PTA observing run will acquire 30 scans
on each pulsar at each frequency.  Each scan yields a single TOA, so one begins with 30 TOAs per frequency per day.
For two reasons, one typically compiles these 30 successive scans into a single ``daily averaged" TOA.  First, the averaging
improves the signal-to-noise ratio of the TOAs by roughly a factor of $\sqrt{30}$.  Second, pulsar timing arrays (PTAs)
are not as sensitive to gravitational waves from either massive black-hole binaries or relic sources of 
gravitational waves sources in the vicinity of frequencies near 1/min as they are to sources near 1/year because of
the expected spectrum of those sources (see
articles in this volume by Sesana, Cornish, Siemens, Cordes and Ellis), so there is little need to maintain
the time resolution of the array near the minute-long timescales.

There are two ways to accomplish this daily averaging, and both depend on having a good timing model for
the pulsar.  First, one can average the profiles together into a single
profile and compute a single TOA.  Alternatively one can average the TOAs together.
Averaging TOAs together is non-trivial, but straightforward.
First, the timing model is used to create residuals from each of the TOAs.  One then
performs a weighted average of the residuals, yielding a single residual and a single uncertainty.  This residual
is then added to the model prediction of an arrival time in the center of the time-range represented by the
original TOAs. 

The result is a new set of TOAs, one for each day for each frequency at which the pulsar was observed.   This
resulting set of TOAs is generally what is used for the remainder of the analysis.

This procedure could in principle be carried even farther in order to combine the TOAs at widely different frequencies
into a single TOA representing an ``infinite frequency" (non-ISM-disturbed) TOA\citep{Keith13}. 
The procedure would be similar to that described above. (See \S \ref{sec:largebandwidth} for another way to average timing
data over large ranges in frequency.)  In practice, this is not common yet.  Typically
pulsar astronomers like to look at the nuances in the residuals at different frequencies,
and some frequencies may be missing on some days.  All these caveats make
astronomers reluctant to absorb all that detailed information into a single TOA.  

\section{When things go wrong with determining TOAs}
\label{sec:wrong}

\subsection{Polarization}
As described above, when the data profile shape does not match the
template profile shape, biases are introduced in the TOAs.  While a
constant error of this sort is unimportant, profile shape errors that
vary as a function of time and/or radio frequency can corrupt timing
solutions.  One important source of such errors is the polarization
properties of the pulsar signal.  Pulsar radiation is generally highly
polarized, and the polarization state changes across the pulse profile
(e.g. Stairs (2001))
\nocite{Stairs01}.  The response of the radio telescope receiver
and backend systems can be represented by a matrix transformation (the
Mueller matrix) that connects the ``true'' astronomical source
parameters with the observed values. In order to calibrate out the
instrumental response this matrix typically 
must be determined empirically from observations of astronomical calibration sources 
(e.g., \cite{vanStraten04}).   Errors or
uncertainties in this process alter the profile shapes and affect timing
results.  One proposed solution to this problem is to determine TOAs
using the ``invariant interval'' -- this quantity, related to the
unpolarized portion of the signal, is insensitive to calibration errors
\citep{Britton00}.  However this method has only proved useful for a
small number of pulsars.  A second approach, called matrix template matching, uses the full polarization
content of the signal while simultaneously determining TOAs
\citep{vanStraten06}.  This also reduces sensitivity to calibration
errors and in some cases also improves TOA accuracy.  Recent work has
shown how pulsars can themselves be used as polarization calibrators
(\cite{vanStraten13}), allowing calibration solutions to be determined
from the brightest sources in a dataset, then applied to the rest.

\subsection{The large-bandwidth problem and its possible solutions}
\label{sec:largebandwidth}

\subsubsection{The problem}

The large-bandwidth problem stems from our inability to identify a fiducial point in the pulse profile.  A TOA is not
so much the arrival time of a pulse as it is the arrival time of a particular template.  One can use the leading edge
of a particular template as the fiducial point, the tallest bin in the template, or the phase of the 1st Fourier
component of the template.  As long as one is consistent about the choice, and uses the same template for the whole data set 
then generally the choice does not matter.

\begin{figure}
\includegraphics[width=.5\textwidth]{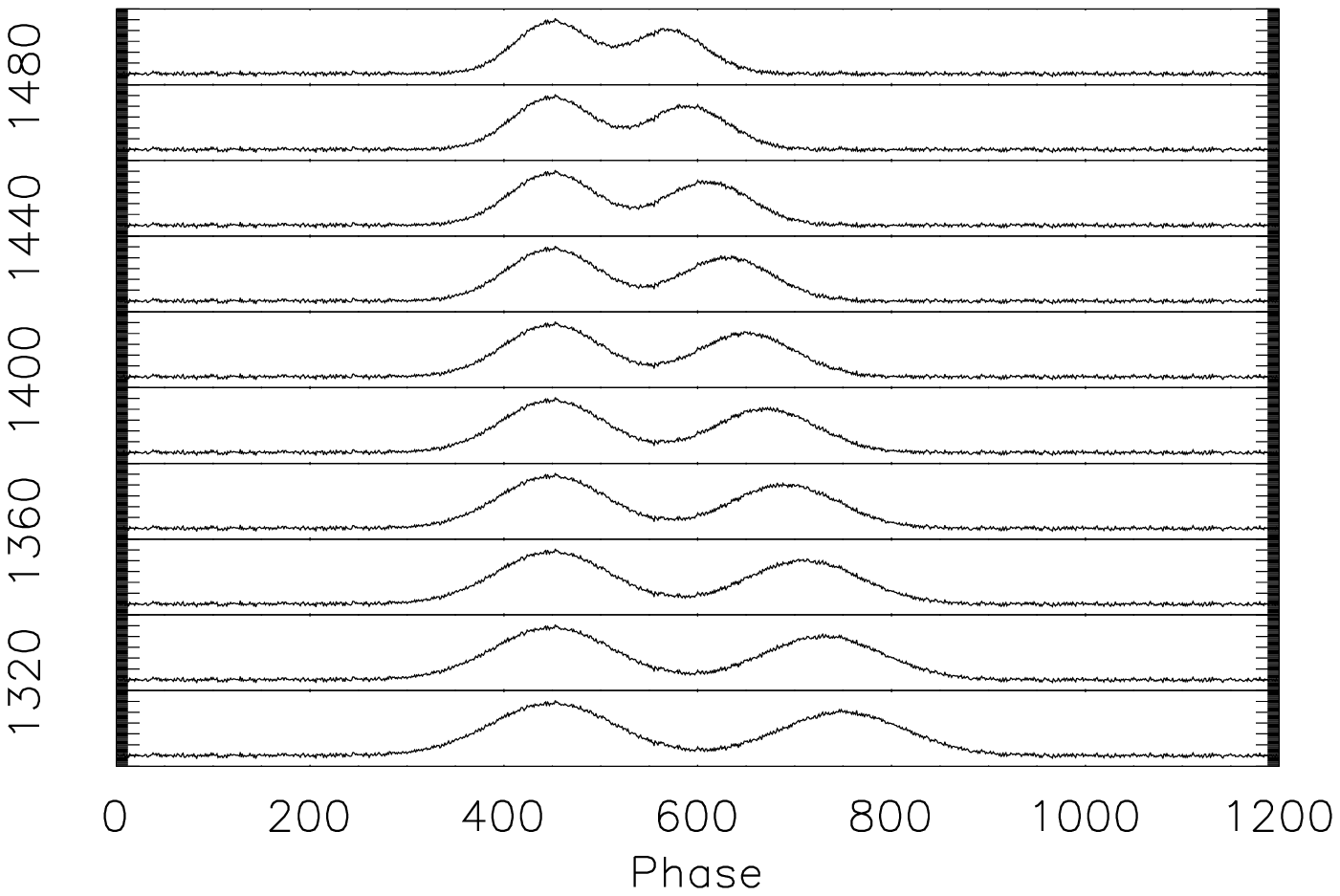}
\includegraphics[width=.5\textwidth]{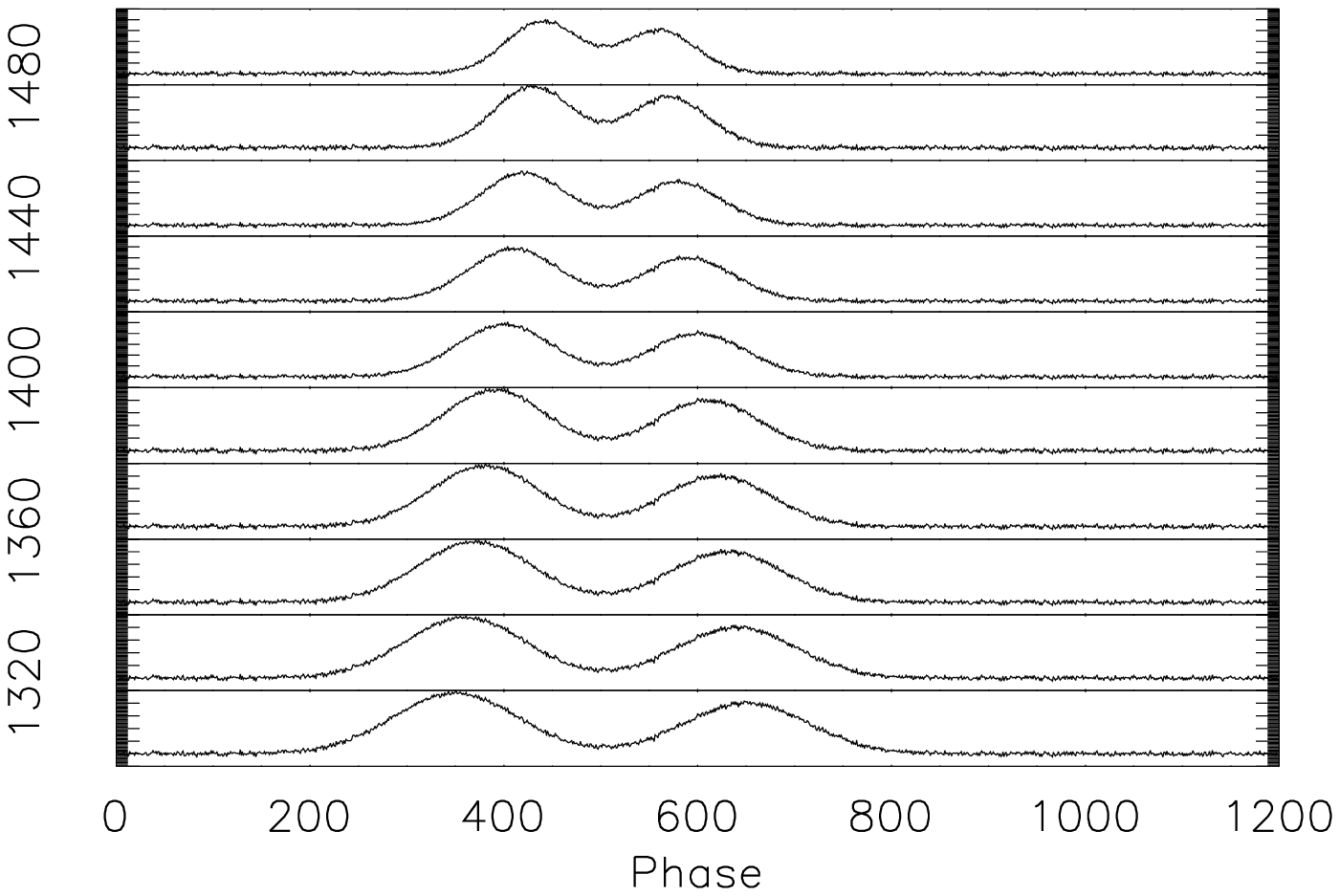}
\caption{A simulated pulsar showing evolution of the pulse profile with frequency, and two different possible alignments
of the pulse profile over frequency.
\label{fig:evolution}
}
\end{figure}

However, pulse profiles evolve over frequency, i.e. they have a different shape at low
frequency than they do at high frequency, and they change continuously in between.  So first consider the problem of
using just two frequencies, eg. 1400 and 2400 MHz.  The pulse profile looks slightly different at each, perhaps 
the profile is double-peaked and the relative height ratio between the two peaks is different at the two frequencies.  
Figure \ref{fig:evolution} shows how a pulsar might evolve over frequency.  
A close match between the shape of the template and the shape of the profile is crucial for reliable timing, so it
would be ill advised to use the same template for both frequencies, and one must use one template for 1400 and a
different one for 2400. 

But now the choice of fiducial point is problematic.  One has insufficient information to choose a fiducial point
that represents a physical location on the star.  In the case of a single-sharp-peaked profile the choice would be
likely be the peak, but what part of the peak would be chosen?  The peak bin?  The leading half-max point?  Most profiles
are much more complicated than a single sharp peak, and the choices are infinite.  In the case of two-frequency data
any mistake is absolved by the fit to the column density of electrons described above.  In other words, a 
bad choice of fiducial
point will amount to a constant offset between the 1400 MHz and 2400 MHz data.  The constant offset will be absorbed
by the fit to electron column density (usually called dispersion measure or DM) by slightly changing the fitted DM.
As a matter of nuance, we should also say that often, especially in precision timing work, the DM is allowed to
change over the span of the data set. Physically this corresponds to 
changes in the electron content along the pulsar-observatory line-of-sight due to the motion of the
pulsar, Earth, and ISM.
The basic idea, however, of a single offset DM absorbing any error in fiducial point designation holds true. 

Now imagine that instead of 2 frequencies there are 16.  This was the case for early North American
Nanohertz Observatory of Gravitational Waves (NANOGrav) data which
consisted of 16 4-MHz channels (64 MHz total bandwidth).  Now with the use of
more recent  backends (e.g. GUPPI, PUPPI, see \cite{GUPPI08}) 
NANOGrav data is typically 512 1.5-MHz 
channels (800 MHz total).
Each of the 16 templates looks slightly different, and the choice of fiducial point
is not obvious.  Because there are 16 frequencies instead of just 2, the DM fit no longer absolves
one of confronting the issue.  What should one do?  This is the large-bandwidth problem.  Following are descriptions of
solutions to this problem.

\subsubsection{The solutions}
\label{sec:possible_alternates}

The solution in \citet{Demorest13} is inspired by an extrapolation of the two-frequency method described above.
One chooses a fiducial point in each of the 32 templates, and as above, the choice does not matter. One 
generates TOAs for each profile, and applies TEMPO to the TOAs with the following addition:  an additional
fittable arbitrary offset (known in TEMPO as a ``jump") is added for 31 of the 32 frequencies.   In other words,
one is pleading ignorance as to the alignment of the 32 templates, and allows TEMPO to find the alignment
that best reduces the RMS.  The 31 jumps are tantamount to the DM fit in the two-frequency case.  In both cases,
the fit allows adjustment for any fiducial-point errors made in the use of the templates.  The analysis is
concluded by averaging the residuals results together on any given day. \cite{Demorest13} demonstrated that this
successfully removes the ISM contribution from the data.  Criticisms include a concern that the extra free
parameters may absorb some of the signal of interest.

Another solution that shows promise, but has not yet been placed into practice is
to create a single
two-dimensional template (intensity vs pulse phase vs frequency)  
rather than creating a one-dimensional template (intensity vs pulse phase) at a number
of different frequencies.
The phase of the template is adjustable smoothly as a function of frequency, but only adjustable
perhaps according to the cold-plasma dispersion relation that governs the arrival time of pulses as a function
of frequency.  A model like this has been demonstrated to work on simulated data using Bayesian analysis
by \citet{Messenger11} and is also being developed by both \cite{Pennucci13} and \cite{Stappers13} using
$\chi^2$ minimization to optimize the solution.  This scheme has the advantage of elegance,
i.e. one only needs a single template for a single data scan no matter how wide the bandwidth of the scan.

\subsection{Pulse-shape Variations}
There remain issues of pulse-shape changes 
even in {\em single} frequency data (i.e. apart from the wide bandth issues we discuss above).  
Some observed changed are thought to be intrinsic to the pulsar
\citep{Kramer99_1022} and some are calibration errors \citep{Britton00}.
It is possible to develop a completely general approach to the problem
of pulse shape variation and timing errors.  Rather than trying to
anticipate the various causes of profile shape variations, these can be
characterized empirically in the data using principal components
analysis (PCA; \cite{Demorest07, Oslowski11}).  In this approach the
observed profile variations in the residual (post template-matching)
profiles are measured via PCA.  Correlations between the profile
variations and timing residuals are then measured and used to correct
the TOAs.  A recent extension of this procedure that uses full
polarization profiles has shown an 40\% improvement in the timing of
J0437$-$4715 (\cite{Oslowski13}).

\section{Calibrating telescope offsets using pulsars}
Finally, the difference in telescope hardware and software causes observing-system-dependent delays in TOAs from any 
particular machine.  If only one observing system is used to create a set of TOAs this delay is irrelevant.  However,
in this era of multi-telescope multi-backend observing the offsets between machines are important.  The most common
practice is to fit an arbitrary offset in phase between TOAs from different observing systems.  For example, in a
set of TOAs employing two different backends, an offset would be required between the two systems for each pulsar at
each frequency \citep{Verbiest08}.  These extra fitted parameters can serve to absorb gravitational wave signal.
Rather than having to determine these instrumental offsets from pulsar data, \cite{Manchester13} developed a system to calibrate out the offsets by injecting a locally generated pulsed signal into the signal path.
 
\section{Not included in this article}    
Readers may be interested in some of the following ideas not discussed in this article.
First, Dan Stinebring and the Interstellar Medium Mitigation group of NANOGrav are mounting an effort
to solve for the ISM scattering function using cyclic spectroscopy, a technique brought to pulsar timing by \citet{Demorest11}.  This work shows great promise and is the
subject of an article by Stinebring in this volume.    Second, a number of mode-changing regular (not millisecond)
pulsars have been identified in which the pulsar switches between two different period derivatives \citep{Lyne10}.  In the
absence of the identification of such a mode, an observer might think that the pulsar was exhibiting simple
stochastic timing noise.  In fact, synchronously with the mode switching, the pulsar's profile actually changes
shape, so there is some promise that the analysis could properly and prescriptively account for the two modes.
This mode-switching behavior has not been shown to exist in MSPs, but it has not been ruled out.

\ack
The authors acknowledge useful conversations with Brian Christy and the NANOGrav timing group.
A.N.L. gratefully acknowledges the support of National Science Foundation grants PIRE OIP 09-68296  and AST CAREER 07-48580.
The National Radio Astronomy Observatory is a facility of the National Science Foundation operated under cooperative agreement by Associated Universities, Inc.


\end{document}